\documentstyle[11pt]{article}
\newcommand{\beq}{\begin{equation}}
\newcommand{\eeq}{\end{equation}}
\newcommand{\bdm}{\begin{displaymath}}
\newcommand{\edm}{\end{displaymath}}
\newcommand{\beqr}{\begin{eqnarray}}
\newcommand{\eeqr}{\end{eqnarray}}
\newcommand{\beqrn}{\begin{eqnarray*}}
\newcommand{\eeqrn}{\end{eqnarray*}}

\begin{document}
\title{Some results on the eigenfunctions of the quantum trigonometric 
Calogero-Sutherland model related to the Lie algebra $D_4$}
\author{J. Fern\'andez N\'u\~{n}ez$^{\;\dagger}$, W. Garc\'{\i}a Fuertes$^{\;
\dagger}$,  A.M. Perelomov$^{\;\ddagger\;}$
\footnote{On leave of absence from the Institute for Theoretical 
and Experimental Physics, 117259, Moscow, Russia. 
Current E-mail address: perelomo@mpim-bonn.mpg.de}\\
{\normalsize {\it $^\dagger$ Departamento de F\'{\i}sica, 
Facultad de Ciencias, Universidad de Oviedo, E-33007 Oviedo, Spain}} \\
{\normalsize {\it $^\ddagger$ Max Planck Institut f\"{u}r Mathematik, 
D-53111 Bonn, Germany}}}
\date{}
\maketitle
\begin{abstract}\noindent
We express the Hamiltonian of the quantum trigonometric Calogero-Sutherland 
model related to the Lie algebra $D_4$ in terms of a set of Weyl-invariant 
variables, namely, the characters of the fundamental representations 
of the Lie algebra. This parametrization allows us to solve for the energy 
eigenfunctions of the theory and to study properties of the system 
of orthogonal polynomials associated to them such as recurrence relations 
and generating functions.
\end{abstract}
\section{Introduction}
Integrable models play a prominent role in theoretical physics. 
The reason is not only the direct phenomenological interest of some 
of them, but also the fact that they often provide some deep insights 
into the mathematical structure of the theories in which they arise. 
Sometimes, they even reveal unexpected relations among different physical 
or mathematical theories. In classical mechanics, integrability not only 
shows up itself in some of the most important and time-honored problems, 
such as the Keplerian motion or the Lagrange or Kovalevskaya top. 
It appears also in a plethora of new hypothetical, highly nontrivial sytems 
discovered mainly during the three last decades of the past century 
(see \cite{ca01, pe90} for comprehensives reviews). Among these, 
the so-called Calogero-Sutherland  models form a distinguised class. 
The first analysis of a system of this kind was performed by Calogero 
\cite{ca71} who studied, from the quantum standpoint, the dynamics on the 
infinite line of a set of particles interacting pairwise by rational 
plus quadratic potentials, and found that the problem was exactly 
solvable. Soon afterwards, 
Sutherland \cite{su72} arrived to similar results for the quantum problem 
on the circle, this time with trigonometric interaction, 
and Moser \cite{mo75} showed that the classical version of both models 
enjoyed integrability in the Liouville sense. The identification of the 
general scope of these discoveries comes with the work 
of Olshanetsky and Perelomov \cite{op77,op78}, who realized that it 
was possible to associate models of this kind to all the root sytems 
of the simple Lie algebras, and that all these models were integrable, 
both in the classical and in the quantum framework \cite{op81,op83}. 
Nowadays, there is a widespread interest in this type of integrable 
systems, and many mathematical and physical applications for them 
have been found, see for instance \cite{dv00}. 

The eigenfunctions of the Calogero-Sutherland Hamiltonian associated 
to the root system of a simple Lie algebra $L$ are proportional 
to some polynomials which form a complete orthogonal system in 
the quantum Hilbert space. For the specials values $\kappa _\alpha =1$, 
where $g_\alpha =\kappa _\alpha (\kappa _\alpha -1)$ are the coupling 
constants, they coincide with the irreducible characters of $L$. 
For $L=A_n$, these polynomials provide natural generalizations 
to $n$ variables of the classical orthogonal polynomials in one 
indeterminate. In particular, for the case with a trigonometric potential, 
one obtains a generalized system of Gegenbauer polynomials. 
As it was shown in the papers \cite{pe98a,prz98,pe99}, these generalized 
Gegenbauer polynomials obey a set of recurrence relations which constitute 
a $\kappa$-deformation of the Clebsch-Gordan series of the algebra. 
The finding of these recurrence relations opened the way to obtain many 
concrete results on the system of polynomials, as for example explicit 
expressions, ladder operators or generating functions \cite{flp01,fp02}. 
The recurrence relations are also the key ingredient to formulate 
a perturbative approach to most general among the Calogero-Sutherland 
models, that involving the Weierstrass $\wp$-function as potential 
\cite{ffp03}. 

The aim of this paper is to extend some of the results which have been 
obtained for $A_n$ to the polynomials related to other simple algebras. 
We think that it is a good idea to begin with a concrete case. 
We choose to work in the first place the problem associated 
to $D_4$ because of the triality symmetry exhibited by this algebra, 
which will help us in simplifying of the treatment. 
The organization of the paper is as follows. In Sect. 2, we explain 
how to express the Calogero-Sutherland Hamiltonian in terms of 
the fundamental characters of the algebra and how to solve the 
Schr\"odinger equation. Then, in Sect. 3, we obtain the main recurrence 
relations among the polynomials and use them to give algorithms 
to calculate some subsets of them. Sect. 4 is devoted to find the 
generating functions for some classes of characters and monomials 
functions of $D_4$. More recurrrence relations and some other relevant 
results are included in Sect. 5, and finally, in Sect. 6, we give some 
brief conclusions. 
Also, we offer two appendices. In Appendix A, for the convenience of 
the reader, we collect some of the basic facts about $D_4$ 
which we use in the main text. In Appendix B we list some polynomials, 
characters and monomial functions.

\section{The eigenvalue problem}
The Hamiltonian operator for the trigonometric Calogero-Sutherland model 
related to the root system of a simple Lie algebra of rank $r$ has the form
\beq 
H=\frac{1}{2}(p,p)+\sum_{\alpha\in R^+}\kappa_\alpha
(\kappa_\alpha-1)\sin^{-2}(\alpha,q),
\eeq where $q=(q_1,\ldots,q_r)$, $p=(p_1,\ldots,p_r)$, $(\ ,\ )$ is the usual 
euclidean inner product in ${\bf R}^r$, $R^+$ is the set of positive roots 
of the algebra, and $\kappa_\alpha$ are constants such that $\kappa_\alpha=\kappa_\beta$ 
if $\mid\mid\alpha\mid\mid=\mid\mid\beta\mid\mid$. In particular, 
for the case of the algebra $D_4$ (see Appendix A), this leads to 
the following Schr\"odinger equation:
\beqr 
H\Psi^\kappa&=&E(\kappa)\,\Psi^\kappa ,\nonumber\\
H=-\frac{1}{2}\,\Delta&+&\kappa (\kappa-1)\left(\sum_{j<k}^4\sin^{-2}
(q_j-q_k)+\sum_{j<k}^4\sin^{-2}(q_j+q_k)\right),\ \ \ 
\Delta=\sum_{j=1}^4\frac{\partial^2}{\partial q_j^2}\,.\label{sch1}
\eeqr The $q$ coordinates are assumed to take values in the $[0,\pi]$ interval, 
and therefore the equation can be interpreted as describing the dynamics 
of a system of four particles moving on the circle. Let us notice that 
there is not translational invariance. We recapitulate some important 
facts about this model which follow from the general structure of the 
quantum Calogero-Sutherland models related to Lie algebras \cite{op83}. 
The ground state energy and the (non-normalized) wavefunction are\beqr 
E_0(\kappa)&=&2 (\rho, \rho)\,\kappa^2= 28\,\kappa^2, \nonumber\\
\Psi_0^\kappa(q)&=&\left\{ \prod_{j<k}^4\sin(q_j-q_k)\sin(q_j+q_k)
\right\} ^\kappa,
\eeqr where $\rho$ is the standard Weyl vector, $\rho=\frac{1}{2}
\sum_{\alpha\in R^+} \alpha$, 
with the sum extended over all the positive roots of $D_4$. 
The excited states depend on a four-tuple of quantum numbers 
${\bf m}=(m_1,m_2,m_3,m_4)$\beqr 
H\,\Psi^\kappa_{\bf m}&=&E_{\bf m}(\kappa)\,\Psi_{\bf m}^\kappa ,\nonumber\\
E_{\bf m}(\kappa)&=&2 (\lambda+\kappa\rho,\lambda+\kappa\rho),\label{105}
\eeqr where $\lambda$ is the highest weight of the irreducible representation 
of $D_4$ labelled by ${\bf m}$, i.e., $\lambda=\sum_{i=1}^4 m_i \lambda_i$, 
and $\lambda_i$ are the fundamental weights of $D_4$. By substitution 
in (\ref{105}) of\beq 
\Psi_{\bf m}^\kappa(q)=\Psi_0^\kappa(q)\,\Phi_{\bf m}^\kappa(q),\label{107}
\eeq 
we are led to the eigenvalue problem
\beq 
-\Delta^\kappa\Phi_{\bf m}^\kappa=\varepsilon_{\bf m}(\kappa)\,
\Phi_{\bf m}^\kappa\label{sch}
\eeq with\beq \Delta^\kappa=\frac{1}{2}\Delta+\kappa\sum_{j<k}^4 \left({\rm ctg}(q_j-q_k)
\left( \frac{\partial}{\partial q_j}-\frac{\partial}{\partial q_k}\right) 
+{\rm ctg}(q_j+q_k)\left( \frac{\partial}{\partial q_j}+\frac{\partial}
{\partial q_k}\right) \right) 
\label{4b} \eeq and\beq \varepsilon_{\bf m}(\kappa)=E_{\bf m}(\kappa)-E_0(\kappa)= 2(\lambda, 
\lambda+2\kappa\rho).\eeq Introducing the inverse Cartan matrix $A_{jk}^{-1}= (\lambda_j,\lambda_k)$, 
we can give a more explicit expression for $\varepsilon_{\bf m}(\kappa)$:\beqr \varepsilon_{\bf m}(\kappa)&=&2\sum_{j,k=1}^4 A_{jk}^{-1}\,m_j m_k+
4\kappa\sum_{j,k=1}^4 A_{jk}^{-1}\,m_j=2(m_1^2+m_3^2+m_4^2)+4m_2^2+
2(m_1 m_3+m_1 m_4+m_3 m_4)\nonumber\\&+&4m_2(m_1+m_3+m_4)+
12\kappa(m_1+m_3+m_4)+20\kappa m_2.\label{eigenvalues}\eeqr 

The main problem is to solve equation (\ref{sch}). As it has been shown for 
the case of the algebra $A_n$ \cite{pe98a,prz98,pe99}, the best way to do that is to use a set 
of independent variables which are invariant under the Weyl symmetry of 
the Hamiltonian, namely, the characters of the four fundamental 
representations of the algebra $D_4$. Unfortunately, the expression of 
these characters in terms of the $q$-variables (which play the role 
of coordinates on the maximal torus of $D_4$) is not very simple. 
Denoting the character of the irreducible representation of maximal 
weight $\lambda_j$ as $z_j$, we find 

\beqrn
z_1&=&\sum_{j=1}^4 x_j+\sum_{j=1}^4 x_j^{-1}\\
z_2&=& \sum_{i<j}^4 x_i x_j +\sum_{i<j}^4 (x_i x_j)^{-1}+\sum_{i,j}^4
x_i^{-1} x_j\\
z_3&=&\bar{x}\sum_{i=1}^4\frac{1}{x_i}+\frac{1}{\bar{x}}
\sum_{i=1}^4x_i,\label{zetas}\\
z_4&=&\bar{x}+\frac{1}{\bar{x}}+\frac{1}{\bar{x}}
\sum_{i<j}^4x_i x_j\\
\eeqrn
where $x_j=e^{2 i q_j}$, and $\bar{x}=\sqrt{x_1 x_2 x_3 x_4}$. 
These expressions make the direct change of variables from $q_i$ to $z_k$ 
quite cumbersome. We refrain from trying that approach, and choose an 
indirect route which has the added advantage of being also applicable 
to other algebras in which the expressions for the characters are even 
more involved. We can infer from (\ref{4b}) the structure 
of $\Delta^\kappa$ when written in the $z$-variables:\beq \Delta^\kappa=\sum_{j,k=1}^4 a_{jk}(z_i)\,\partial_{z_j}\,\partial_{z_k}+
\sum_{j=1}^4 \left[ b_j^{(0)}(z_i)+\kappa b_j^{(1)}(z_i)\right] 
\partial_{z_j}. \label{structure}\eeq On the other hand, as it is well known \cite{ma95}, the 
$\Phi_{\bf m}^\kappa$ are polynomials which, with some precise partial 
ordering for the monomials to be described later, start as follows:\beq \Phi_{\bf m}^\kappa(z_i)=P_{\bf m}^\kappa(z_i)=z_1^{m_1}z_2^{m_2}z_3^{m_3}
z_{4}^{m_4}+\cdots .\eeq Therefore, making use of (\ref{eigenvalues}), we conclude that\beqr a_{jk}(z_i)&=&2\,A_{jk}^{-1}\,z_j z_k+{\rm lower\ order\ terms},\nonumber\\b_j^{(r)}(z_i)&=&c_j^{(r)} z_j+d_j^{(r)},\ \ \ \ \ r=0,1.\label{101}\eeqr 

Now, to obtain the full expressions for these coefficients, we rely on 
the fact that, for $\kappa=1$, the $P_{\bf m}^\kappa$ polynomial gives 
the character of the irreducible representation of $D_4$ with maximal weight 
$\sum_{i=1}^4m_i\lambda_i$, while for $\kappa=0$ the same polynomial is 
the corresponding symmetric monomial function \cite{op83}. Both, characters and monomial 
functions, can be computed by using  the information available in the 
literature (see, for instance, the ``Reference Chapter" of \cite{ov90}). 
In fact, the following short list of polynomials \beqrn P_{2,0,0,0}^{(1)} (z) &=&z_1^2-z_2-1,\\P_{1,1,0,0}^{(1)}(z) &=&z_1 z_2-z_3 z_4, \\P_{1,0,1,0}^{(1)}(z) &=&z_1 z_3- z_4, \\P_{0,2,0,0}^{(1)}(z) &=&z_2^2-z_1 z_3 z_4+z_2, \\P_{2,0,0,0}^{(0)}(z) &=&z_1^2-2 z_2\eeqrn is all we need to obtain $\Delta^\kappa$. By substituting  these 
polynomials in (\ref{sch}) and using (\ref{eigenvalues}), (\ref{structure}), 
(\ref{101}) and the triality symmetry (that here implies that the final 
expression for $\Delta^\kappa$ should be invariant under permutations 
of the indices 1,3,4), we get enough simple linear algebraic equations 
to  fix all the coefficients. We give here only the final result:\beqr \frac{1}{2}\Delta^\kappa&=&\left( z_1^2-2z_2-8\right) \partial_{z_1}^2+
\left[ 2\,z_2^2-4\left( z_1^2+z_3^2+z_4^2\right) -2\,z_1\,z_3\,z_4+8\,
z_2\right] \partial_{z_2}^2+\left( z_3^2-2z_2-8\right) \partial_{z_3}^2
\nonumber\\
&+&\left( z_4^2-2z_2-8\right) \partial_{z_4}^2+(2z_1z_2-6z_3z_4- 8z_1)\,
\partial_{z_1}\,\partial_{z_2}+(z_1\,z_3-8\,z_4)\,\partial_{z_1}\,
\partial_{z_3}\nonumber\\
&+&(z_1z_4-8z_3)\,\partial_{z_1}\,\partial_{z_4}+(2z_2z_3-6z_1z_4- 
8z_3)\,\partial_{z_2}\,\partial_{z_3}+(2z_2z_4-6z_1z_3-8z_4)\,
\partial_{z_2}\,\partial_{z_4}\nonumber\\
&+&(z_3z_4-8z_1)\,\partial_{z_3}\,\partial_{z_4}+(6\kappa+1)\,z_1\,
\partial_{z_1}+[2(5\kappa+1)\,z_2+8(\kappa-1)]\,\partial_{z_2}+
(6\kappa+1)\,z_3\,\partial_{z_3}\nonumber\\
&+&(6\kappa+1)\,z_4\,\partial_{z_4} \label{delta}.\eeqr 

Once the explicit expression for the operator $\Delta^\kappa$ in the $z$ 
variables is given, the Schr\"odinger equation can be solved iteratively. 
By direct application of $\Delta^\kappa$ to $z^{\bf m}\equiv z_1^{m_1}\,
z_2^{m_2}\,z_2^{m_3}\,z_4^{m_4}$, we find\beqr \Delta^\kappa z^{\bf m}&=&\varepsilon_{\bf m}(\kappa)\,z^{\bf m}-
\sum_{i=1}^4 a_{\bf m}^i\,z^{{\bf m}-\alpha_i}-\sum_{j\in I}b_{\bf m}^j\,
z^{{\bf m}-(\alpha_2+\alpha_j)}-\sum_{ij\in T}c_{\bf m}^{ij}\,
z^{{\bf m}-(\alpha_2+\alpha_i+\alpha_j)}\nonumber\\
&-&2\,a_{\bf m}^2\sum_{ij\in T}z^{{\bf m}-(2\alpha_2+\alpha_i+\alpha_j)}-
d_{\bf m}\,z^{{\bf m}-(\alpha_1+2\alpha_2+\alpha_3+\alpha_4)}
\nonumber\\
&-&4\sum_{j\in I}a_{\bf m}^j\,z^{{\bf m}-(\alpha_1+2\alpha_2+\alpha_3+
\alpha_4+\alpha_j)}, \label{recu}
\eeqr 
where the sets of indices are $I=\{1,3,4\}$ and $T=\{13,14,34\}$, and\beqrn a_{\bf m}^i&=&4\,m_i(m_i-1),\ \ \ \ \ \ \ \ \ b_{\bf m}^j=12 m_2 m_j,\\c_{\bf m}^{ij}&=&16\,m_i\,m_j,\ \ \ \ \ \ \ \ \ \ \ \ \ \ \ 
d_{\bf m}(\kappa)=16\,m_2\left( 2-m_2-\kappa+\sum_{j\in I} m_j\right) .\eeqrn All monomials in $\Delta^\kappa z^{\bf m}$ take the form $z^{{\bf m}-\mu}$ 
with $\mu$ as a positive root. Thus, the polynomial $P_{\bf m}^\kappa$ 
has the form\beq P_{\bf m}^\kappa(z)=\sum_{\mu \in Q^+({\bf m})} c_\mu z^{{\bf m}-\mu},\eeq where we choose the normalization $c_0=1$ and, if $Q^+$ is the cone of positive roots,
\beq Q^+({\bf m})=\left\{ \mu\in Q^+ \mid z^{{\bf m}-\mu}\ {\rm is\ well\ defined\ if\ }z_1\,z_2\,z_3\,z_4=0\right\}.\eeq The above-mentioned partial ordering of  monomials is given simply by the height of $\mu$, i. e. $z^{{\bf m}-\mu_1}> z^{{\bf m}-\mu_2}$ if $ht(\mu_1)<ht(\mu_2)$. From (\ref{recu}), the coefficients $c_\mu$ obey the 
iterative formula\beq c_\mu=\frac{N_\mu}{\varepsilon_{{\bf m}-\mu}(\kappa)-
\varepsilon_{\bf m}(\kappa)}\eeq with\beqrn N_\mu &=&\sum_{i=1}^4 a_{{\bf m}-(\mu-\alpha_i)}^i\,c_{\mu-\alpha_i}+
\sum_{j\in I}b_{{\bf m}-(\mu-\alpha_2-\alpha_j)}^j\,
c_{\mu-(\alpha_2+\alpha_j)}+\sum_{ij\in T}c_{{\bf m}-(\mu-\alpha_2-\alpha_i-
\alpha_j)}^{ij}\,c_{\mu-(\alpha_2+\alpha_i+\alpha_j)}\\
&+& 2 \sum_{ij\in T} a_{{\bf m}-(\mu-2\alpha_2-\alpha_i-\alpha_j)}^2\,
c_{\mu-(2\alpha_2+\alpha_i+\alpha_j)}+d_{{\bf m}-(\mu-\alpha_1-2\alpha_2-
\alpha_3-\alpha_4)}\,c_{\mu-(\alpha_1+2\alpha_2+\alpha_3+\alpha_4)}\\
&+& 4\sum_{j\in I}a_{{\bf m}-(\mu-\alpha_1-2\alpha_2-\alpha_3-\alpha_4-
\alpha_j)}^j\,c_{\mu-(\alpha_1+2\alpha_2+\alpha_3+\alpha_4+\alpha_j)}.\eeqrn Along with the explicit expressions for the roots given in Appendix A, 
it is suitable for the implementation on a symbolic computer program. 
A list of polynomials obtained through the use of this formula is offered 
in Appendix B. 

\section{The structure of the recurrence relations}
As it is well known, all the systems of orthogonal polynomials in one 
indeterminate $z$, such that $P_m(z)=z^m+\cdots $ satisfy a recursive 
formula $z\,P_m(z)= a_m\,P_{m+1}(z)+b_{m}\,P_m(z)+c_m\,P_{m-1}(z)$. 
In particular, the orthogonal polynomials associated to the trigonometric 
Calogero-Sutherland model for the case of two particles and Lie algebra 
$A_1$ are the classical Gegenbauer polynomials, whose recursive formula is 
known to be \bdm z\,P_m^\kappa(z)=P_{m+1}^\kappa(z)+\frac{m(m-1+2\kappa)}{(m-1+\kappa)
(m+\kappa)}\,P_{m-1}^\kappa(z).\edm This formula is reminiscent of the Clebsch-Gordan series for $A_1$. 
In fact, for $\kappa=1$ it reduces exactly to this Clebsch-Gordan series: 
the polynomials are the characters of $A_1$ and the coefficents are equal to one. 
Immediately the question arises about the existence of analogous 
recurrence relations, i.e., with the structure of $\kappa$-deformations of 
the corresponding Clebsch-Gordan series, for the polynomials related to 
Calogero-Sutherland models associated to other simple Lie algebras. 
As it was shown in \cite{pe98a}, the answer turns out to be in the 
afirmative for all root systems, but to obtain the expressions for the 
deformed coefficients it is necessary to proceed through a case-by-case 
analysis. Once the coefficients are known, many applications are possible. 
The aim of this section is to fix the structure of the basic recurrence 
relations for the case of $D_4$ and to give a simple illustration of 
their use. 

We want to study the formulas for $z_i P_{\bf m}^\kappa(z)$, $i=1,2,3,4$. 
Therefore, as $P_{\bf m}^{(1)}(z)=z_i$ for $m_j=(\delta_{ji})$, and the 
recursive formulas are deformations of the Clebsch-Gordan series, we need to 
know the weights of the irreducible representations whose integral dominant 
weights are $\lambda_1$, $\lambda_2$, $\lambda_3$ and $\lambda_4$. 
For the case of $\lambda_1$, $\lambda_3$ and $\lambda_4$, these 
representations have dimension eight. On the other hand, if we act on 
the highest weight with the Weyl group in the way explained in the 
Appendix A, we obtain eight different weights. Thus, these representations 
include only one orbit of the Weyl group and we are done. For the case of 
$\lambda_2$, the representation has dimension  28 and the orbit of the Weyl 
group containing $\lambda_2$ has only 24 elements. But $\lambda_2=
\alpha_{12}^+$, the highest root, and thus this representation is the adjoint 
one and includes a second orbit: the Cartan subalgebra, with four elements 
of weight zero. Let us summarize. 
\begin{itemize}\item Weights in $z_1$:$\quad \pm\lambda_1,\quad \pm(\lambda_1-\lambda_2),
\quad \pm(\lambda_2-\lambda_3-\lambda_4),\quad \pm(\lambda_3-\lambda_4).$

\item Weights in $z_2$:$\quad \pm\lambda_2,\quad \pm(\lambda_2-2\lambda_j),
\quad \pm(2\lambda_2-\lambda_1-\lambda_3-\lambda_4),\quad 
\pm(\lambda_2+\lambda_i-\lambda_j-\lambda_k),\quad 
\pm(\lambda_i+\lambda_j-\lambda_k),\quad 
\pm(\lambda_2-\lambda_1-\lambda_3-\lambda_4),\ 0$, with $i,j,k\in I.$

\item Weights in $z_3$:$\quad \pm\lambda_3,\quad \pm(\lambda_3-\lambda_2),
\quad \pm(\lambda_2-\lambda_1-\lambda_4),\quad \pm(\lambda_1-\lambda_4).$

\item Weights in $z_4$:$\quad \pm\lambda_4,\quad \pm(\lambda_4-\lambda_2),
\quad \pm(\lambda_2-\lambda_1-\lambda_3),\quad \pm(\lambda_1-\lambda_3).$
\end{itemize}
With these weights, the structure of the recurrence relations results to be 
as follows: 
\beqrn z_1\,P_{m_1,m_2,m_3,m_4}^\kappa(z)&=& P_{m_1+1,m_2,m_3,m_4}^\kappa(z) +
a_{\bf m}^1(\kappa)\,P_{m_1-1,m_2,m_3,m_4}^\kappa(z) +b_{\bf m}^1(\kappa)\,
P_{m_1+1,m_2-1,m_3,m_4}^\kappa(z) \\
&+& c_{\bf m}^1(\kappa)\,P_{m_1-1,m_2+1,m_3,m_4}^\kappa(z) +d_{\bf m}^1
(\kappa)\,P_{m_1,m_2+1,m_3-1,m_4-1}^\kappa(z)\\
&+&e_{\bf m}^1(\kappa)\,P_{m_1,m_2-1,m_3+1,m_4+1}^\kappa(z) 
+ f_{\bf m}^1(\kappa)\,P_{m_1,m_2,m_3+1,m_4-1}^\kappa(z)\\ 
&+&g_{\bf m}^1(\kappa)\,P_{m_1,m_2,m_3-1,m_4+1}^\kappa(z) \eeqrn 
\beqrn z_2\,P_{m_1,m_2,m_3,m_4}^\kappa(z) &=&P_{m_1,m_2+1,m_3,m_4}^\kappa(z)+
A_{\bf m}(\kappa)\,P_{m_1,m_2-1,m_3,m_4}^\kappa(z)+B_{\bf m}(\kappa)^{1\pm}
\,P_{m_1\pm 2,m_2\mp 1,m_3,m_4}^\kappa(z)\\
&+& B_{\bf m}(\kappa)^{3\pm}\,P_{m_1,m_2\mp 1,m_3\pm 2,m_4}^\kappa(z)+
B_{\bf m}(\kappa)^{4\pm}\,P_{m_1,m_2\mp 1,m_3,m_4\pm 2}^\kappa(z)\\
&+& C_{\bf m}(\kappa)^{\pm}\,P_{m_1\mp 1,m_2\pm 2,m_3\mp 1,m_4\mp 1}^\kappa 
(z)+ D_{\bf m}(\kappa)^{1\pm}\,P_{m_1\pm 1,m_2\pm 1,m_3\mp 1,m_4\mp 1}^
\kappa(z)\\
&+& D_{\bf m}(\kappa)^{3\pm}\,P_{m_1\mp 1,m_2\pm 1,m_3\pm 1,m_4\mp 1}^
\kappa(z)+ D_{\bf m}(\kappa)^{4\pm}\,P_{m_1\mp 1,m_2\pm 1,m_3\mp 1,m_4\pm 1}^
\kappa(z)\\
&+& E_{\bf m}(\kappa)^{1\pm}\,P_{m_1\mp 1,m_2,m_3\pm 1,m_4\pm 1}^\kappa(z)+
E_{\bf m}(\kappa)^{3\pm}\,P_{m_1\pm 1,m_2,m_3\mp 1,m_4\pm 1}^\kappa(z)\\
&+& E_{\bf m}(\kappa)^{4\pm}\,P_{m_1\pm 1,m_2,m_3\pm 1,m_4\mp 1}^\kappa(z)+
F_{\bf m}(\kappa)^{\pm}\,P_{m_1\pm 1,m_2\mp 1,m_3\pm 1,m_4\pm 1}^\kappa(z)\\
&+& G_{\bf m}(\kappa)\,P_{m_1,m_2,m_3,m_4}^\kappa(z),
\eeqrn 
\beqrn z_3\,P_{m_1,m_2,m_3,m_4}^\kappa(z) &=&P_{m_1,m_2,m_3+1,m_4}^\kappa(z) +
a_{\bf m}^3(\kappa)\,P_{m_1,m_2,m_3-1,m_4}^\kappa(z) +b_{\bf m}^3(\kappa)\,
P_{m_1,m_2-1,m_3+1,m_4}^\kappa(z) \\
&+& c_{\bf m}^3(\kappa)\,P_{m_1,m_2+1,m_3-1,m_4}^\kappa(z) +
d_{\bf m}^3(\kappa)\,P_{m_1-1,m_2+1,m_3,m_4-1}^\kappa(z) \\
&+&e_{\bf m}^3(\kappa)\,P_{m_1+1,m_2-1,m_3,m_4+1}^\kappa(z) 
+ f_{\bf m}^3(\kappa)\,P_{m_1+1,m_2,m_3,m_4-1}^\kappa(z)\\
&+&g_{\bf m}^3(\kappa)\,P_{m_1-1,m_2,m_3,m_4+1}^\kappa(z) ,\eeqrn 
\beqrn z_4\,P_{m_1,m_2,m_3,m_4}^\kappa(z) &=&P_{m_1,m_2,m_3,m_4+1}^\kappa(z) +
a_{\bf m}^4(\kappa)\,P_{m_1,m_2,m_3,m_4-1}^\kappa(z) +b_{\bf m}^4(\kappa)\,
P_{m_1,m_2-1,m_3,m_4+1}^\kappa(z) \\
&+& c_{\bf m}^4(\kappa)\,P_{m_1,m_2+1,m_3,m_4-1}^\kappa(z) +
d_{\bf m}^4(\kappa)\,P_{m_1-1,m_2+1,m_3-1,m_4}^\kappa(z)\\
&+&e_{\bf m}^4(\kappa)\,P_{m_1+1,m_2-1,m_3+1,m_4}^\kappa(z) 
+f_{\bf m}^4(\kappa)\,P_{m_1-1,m_2,m_3+1,m_4}^\kappa(z) \\
&+&g_{\bf m}^4(\kappa)\,P_{m_1+1,m_2,m_3-1,m_4}^\kappa(z),\eeqrn 
where $B_{\bf m}(\kappa)^{1\pm}\,P_{m_1\pm 2,m_2\mp 1,m_3,m_4}^\kappa(z)$ 
means $B_{\bf m}(\kappa)^{1+}\,P_{m_1+ 2,m_2- 1,m_3,m_4}^\kappa(z)+
B_{\bf m}(\kappa)^{1-}\,P_{m_1- 2,m_2+ 1,m_3,m_4}^\kappa(z)$, etc, and it is understood that all polynomials involving negative quantum numbers are zero. 
The recurrence relations reflect triality in the fact that not all the coefficients 
appearing in these formulas are independent. There are 
coincidences upon permutations of the quantum numbers,
for instance
\beq a^1_{m_1,m_2,m_3,m_4}=a^3_{m_3,m_2,m_1,m_4}=a^4_{m_4,m_2,m_3,m_1},\eeq and similarly for $b_{\bf m}^j, c_{\bf m}^j, d_{\bf m}^j, e_{\bf m}^j, 
f_{\bf m}^j, g_{\bf m}^j$. In the same fashion, we have also\beq B^{1\pm}_{m_1,m_2,m_3,m_4}=B^{3\pm}_{m_3,m_2,m_1,m_4}=B^{4\pm}_
{m_4,m_2,m_3,m_1}\eeq and similarly for $D_{\bf m}^{j\pm}, E_{\bf m}^{j\pm}$.
 
As an example, let us consider a simple case in which only one of the 
quantum numbers is nonvanishing, namely,\beq z_1\,P_{m,0,0,0}^\kappa(z) =P_{m+1,0,0,0}^\kappa(z) +a_m(\kappa)\,
P_{m-1,0,0,0}^\kappa(z) +c_m(\kappa)\,P_{m-1,1,0,0}^\kappa(z) ,\label{rec1}\eeq where we write $a_m(\kappa)=a_{m,0,0,0}^1(\kappa)$ and $c_m(\kappa)=
c_{m,0,0,0}^1(\kappa)$. Using formulae
\beqrn P_{m,0,0,0}^\kappa(z) &=&z_1^m-\frac{m(m-1)\left[ 4\kappa^2+4(m-2)\kappa+
(m-1)(m-2)\right] }{(m-1+\kappa)(m-1+3\kappa)(m-2+\kappa)}\,z_1^{m-2}-
\frac{m(m-1)}{m-1+\kappa}z_1^{m-2}\,z_2+\cdots\\P_{m,1,0,0}^\kappa(z) &=&z_1^m\,z_2+\frac{4\,\kappa(\kappa-1)(m-2+2\kappa)}
{(m+1+5\kappa)(m+2\kappa)(m-1+\kappa)}\,z_1^m+\cdots,\eeqrn we obtain the coeffficients in (\ref{rec1}) \beqrn a_m(\kappa)&=&\frac{m(m+2\kappa)(m-1+4\kappa)(m-1+6\kappa)}{(m-1+\kappa)
(m-1+3\kappa)(m+3\kappa)(m+5\kappa)},\\c_m(\kappa)&=&\frac{m(m-1+2\kappa)}{(m+\kappa)(m-1+\kappa)}.\label{coef1}\eeqrn 
As a byproduct of triality, we can also write other two recurrence relations 
with the same coefficients:
\beqr z_3 P_{0,0,m,0}^\kappa(z) &=&P_{0,0,m+1,0}^\kappa(z) +a_m(\kappa)\,
P_{0,0,m-1,0}^\kappa(z) +c_m(\kappa)\,P_{0,1,m-1,0}^\kappa(z) \nonumber\\z_4\,P_{0,0,0,m}^\kappa(z) &=&P_{0,0,0,m+1}^\kappa(z) +a_m(\kappa)\,
P_{0,0,0,m-1}^\kappa(z) +c_m(\kappa)\,P_{0,1,0,m-1}^\kappa(z) .\eeqr 

The first of these recurrence relations can be used to devise an algorithm 
for the calculation of the polynomials of the form $P_{m,0,0,0}^\kappa(z) $ 
and $P_{m,1,0,0}^\kappa(z) $. By multiplying (\ref{rec1}) by the differential 
operator $\Delta^\kappa-\varepsilon_{m-1,1,0,0}(\kappa)$, the term involving 
$P_{m-1,1,0,0}^\kappa$ cancels. Using the explicit expressions 
(\ref{eigenvalues}), (\ref{delta}), we find
\begin{eqnarray*}  P_{m+1,0,0,0}^\kappa&=&\frac{1}{4(m+\kappa)}\,[\Delta^\kappa,z_1]\,
P_{m,0,0,0}^\kappa(z) -\frac{1+4\kappa}{2(m+\kappa)}\,
z_1 P_{m,0,0,0}^\kappa(z) \\
&+& \frac{m(m+2\kappa)(m-1+4\kappa)(m-1+6\kappa)}{(m-1+\kappa)(m-1+3\kappa)
(m+\kappa)(m+3\kappa)}\,P_{m-1,0,0,0}^\kappa(z) , \end{eqnarray*}
where, from (13),\begin{eqnarray*} 
\left[ \Delta^\kappa,z_1\right] &=& 4\left( z_1^2-2z_2-8\right) 
\partial_{z_1}+2\,(z_1z_3-8z_4)\,\partial_{z_3}+2\,(z_1z_4-8z_3)\,
\partial_{z_4}\\
&+& 4\,(z_1z_2-3z_3z_4-4z_1)\,\partial_{z_2}+2(6\kappa+1)\,z_1. \end{eqnarray*}
where, from (\ref{delta}),\beqrn \left[ \Delta^\kappa,z_1\right] &=& 4\left( z_1^2-2z_2-8\right) 
\partial_{z_1}+2\,(z_1z_3-8z_4)\,\partial_{z_3}+2\,(z_1z_4-8z_3)\,
\partial_{z_4}\\
&+& 4\,(z_2z_2-3z_3z_4-4z_1)\,\partial_{z_2}+2(6\kappa+1)\,z_1.\eeqrn Once the polynomials $P_{m,0,0,0}^\kappa(z) $ are known, the recurrence 
relation $(\ref{rec1})$ provides a formula for each $P_{m,1,0,0}^\kappa(z)$:\beq c_{m+1}(\kappa)\,P_{m,1,0,0}^\kappa(z) =z_1\,P_{m+1,0,0,0}^\kappa(z) -
P_{m+2,0,0,0}^\kappa(z) -a_{m+1}(\kappa)\,P_{m,0,0,0}^\kappa(z) .\eeq 
 
\section{Some generating functions}
We present in this section the generating functions for some characters and 
symmetric monomial functions. Let us consider first the case of the monomial 
functions with only one non-vanishing quantum number in the form 
$P_{m,0,0,0}^{(0)}(z)$. The generating function for this subset is
\beq 
F_0(t,z)=\sum_{m=0}^\infty t^m\,P_{m,0,0,0}^{(0)}(z).
\eeq 
In terms of the $x$ variables, the general expression for these monomial 
functions is\beq P_{m,0,0,0}^{(0)}(x)=\sum_{j=1}^4\left(x_j^m+x_j^{-m}\right),
\label{assumption}\eeq and, in particular, we define $P_{0,0,0,0}^{(0)}(z) =8$. In these variables, 
the computation of $F_0(t,x)$ only requires to sum the geometric series:\beq F_0(t,x)=\sum_{j=1}^4\left(\frac{1}{1-t x_j}+\frac{1}{1-\frac{t}{x_j}}\right).\eeq The change to the original $z$ variables can be done by the inspection of 
the coefficients of the powers of $t$ in both the numerator and denominator 
of this rational expression, with the result\beq F_0(t,z)=\frac{N_0(t,z)}{D(t,z)}, \label{f0}\eeq where\beqr N_0(t,z)&=&8-7z_1\,t+6 z_2\,t^2-5(z_3 z_4-z_1)\,t^3+4(z_3^2+z_4^2-2 z_2-2)
\,t^4-3(z_3 z_4-z_1)\,t^5\nonumber\\
&+&2 z_2\,t^6-z_1\,t^7, \nonumber\\D(t,z)&=&1- z_1\,t+ z_2\,t^2-(z_3 z_4-z_1)\,t^3+(z_3^2+z_4^2-2 z_2-2)\,t^4-
(z_3 z_4-z_1)\,t^5\nonumber\\
&+&z_2\,t^6-z_1\,t^7 +t^8.\eeqr 
There is an alternative approach. As the monomial functions are 
eigenfunctions of $\Delta^{(0)}$ with eigenvalues $\varepsilon_{m,0,0,0}(0)=
2m^2$, we have\bdm \frac{1}{2}\,\Delta^{(0)}\,F_0(t,z)=\sum_{m=0}^\infty m^2t^m\,
P_{m,0,0,0}^{(0)}(z),\edm and, therefore, we can write a differential equation for $F_0(t,z)$:\beq \left[\frac{1}{2} \Delta^{(0)}-(t\;\partial_t)^2\right]F_0(t,z)=
0,\ \ \ \ \ \ F_0(0,z)=8.\eeq One can verify by substitution that (\ref{f0}) satisfies this equation. 
When $F_0(t,z)$ is known, we can easily obtain the generating function\beq G_0(t,z)=\sum_{m=0}^\infty t^m\,P_{m,1,0,0}^{(0)}(z)\eeq by only recalling (\ref{rec1}), which for $\kappa=0$ is simply\beq z_1 P_{m,0,0,0}^{(0)}(z) =P_{m+1,0,0,0}^{(0)}(z) +P_{m-1,0,0,0}^{(0)}(z) +
P_{m-1,1,0,0}^{(0)}(z) .\eeq This gives\beq G_0(t,z)=\frac{M_0(t,z)}{D(t,z)}\eeq with
\begin{eqnarray*} 
M_0(t,z)&=&z_2-4+(6\,z_1-3\,z_3 z_4)\,t\\
&+&(-8-2\,z_1^2-10\,z_2-z_2^2+4\,z_3^2+2\,z_1 z_3 z_4+4\,z_4^2)\,t^2\\
&+&(10\,z_1+5\,z_1z_2-3\,z_1 z_3^2-4\,z_3 z_4+z_2 z_3 z_4-3\,z_1 z_4^2)\,t^3
\\
&+&(8\,z_2-4\,z_1^2+2\,z_2^2-z_2 z_3^2+4\,z_1 z_3 z_4-z_2 z_4^2)\,t^4\\
&+&(-6\,z_1-6\,z_1 z_2-z_3 z_4+z_2 z_3 z_4)\,t^5
+(8+6z_1^2+2\,z_2-z_2^2)\,t^6\\
&+&(-10\,z_1+z_1 z_2)\,t^7+(4-z_2)\,t^8.
\end{eqnarray*}
The computation of the generating functions for the characters 
$P_{m,0,0,0}^{(1)}$ and $P_{m,1,0,0}^{(1)}$ goes through similar arguments. 
In this case, the eigenvalues are $\varepsilon_{m,0,0,0}(1)=2m^2+12m$. Hence, 
\beq F_1(t,z)=\sum_{m=0}^\infty t^m P_{m,0,0,0}^{(1)}(z)\, ,\ \ \ \ \ \ \ \ 
P_{0,0,0,0}^{(1)}(z) \equiv 1\eeq is the solution of the equation\beq \left[ \frac{1}{2}\Delta^{(1)} -(t\,\partial_t)^2-6t\,\partial_t\right] 
F_1(t,z)=0,\ \ \ \ \ \ F_1(0,z)=1.\eeq The Weyl character formula implies that the denominator of $F_1(t,z)$ should 
be the same $D(t,z)$ found before. Thus, we try an Ansatz\beq F_1(t,z)=\frac{N_1(t,z)}{D(t,z)}\eeq and obtain the simple answer\beq N_1(t,z)=1-t^2.\eeq Applying the recurrence relation (\ref{rec1})) we obtain the generating 
function $G_1(t,z)$ for the characters $P_{m,1,0,0}^{(1)}$:
\beq G_1(t,z)=\frac{1}{D(t,z)}\left\{z_2-z_3 z_4 t+(z_3^2+z_4^2-2 z_2-1)\,t^2-
(z_3 z_4-z_1)\,t^3+ z_2\,t^4-z_1\,t^5+t^6\right\} .\eeq 

\section{More recurrence relations and other results}
In this Section, we give the remaining recurrence relations involving 
the product of a fundamental character times a polynomial with only one 
non-vanishing quantum number. We also comment  the existence of some 
peculiar values for $\kappa$ for which the polynomials associated to some 
special excited states are proportional to integer powers of the fundamental 
state wavefunction. 

To obtain the mentioned recurrence relations, it is necessary to compute 
the coefficients of a limited number of terms of the polynomials involved. 
Once the form of these terms is known, we can obtain the coefficients 
in the recurrence relations  solving a system of linear algebraic equations. 
We do not give here the full expressions for the coefficients of the required terms, because 
some of them are too long, and only list them:

\beqrn P_{1,0,m,0}^\kappa(z)&=&z_1 z_3^m+ A\,z_3^{m-1} z_4+\cdots ,\\P_{0,m,0,0}^\kappa(z)&=& z_2^m+ B\,z_2^{m-1}+ C\,z_2^{m-2}+ 
D\,z_1 z_2^{m-2} z_3 z_4+ E\,z_1 z_2^{m-3}z_3 z_4 \\
&+& F\,(z_1^2 z_2^{m-2} +z_2^{m-2} z_3^2+ z_2^{m-2} z_4^2)+\cdots ,\\P_{1,m,0,0}^\kappa(z)&=&z_1 z_2^m+ G\,z_1z_2^{m-1} z_4+H\,z_2^{m-1} z_3 z_4+
\cdots ,\\P_{0,m,1,1}^\kappa(z)&=&z_2^m z_3 z_4+ I\,z_1 z_2^m+\cdots , \\P_{m,0,0,0}^\kappa(z)&=&z_1^m+ J\,z_1^{m-2}+K\,z_1^{m-2} z_2+\cdots ,\\P_{m,1,0,0}^\kappa(z)&=&z_1^m z_2+ L\,z_1^{m-2}z_2+N\,z_1^{m-1} z_3 z_4+ 
M \,z_1^m+\cdots ,\\P_{m,0,1,1}^\kappa(z)&=&z_1^m z_3 z_4+ N\,z_1^{m-1} z_2+ O\,z_1^{m+1}+
\cdots ,\\P_{1,m,1,1}^\kappa(z)&=&z_1 z_2^m z_3 z_4+P\,z_2^m+ Q\,z_1 z_2^{m-1}z_3 z_4+R\,(z_1^2z_2^m+z_2^m z_3^2+z_2^mz_4^2)+S\,z_2^{m+1}+\cdots ,\\P_{2,m,0,0}^\kappa(z)&=&z_1^2 z_2^m+ T\,z_2^m + U\,z_1 z_2^{m-1} z_3 z_4+
W\,z_2^{m+1}+\cdots .\\\eeqrn 

The use of the quantities denoted $A$ to $W$ in the previous formulas in the general structure of the 
recurrence relations give the following results: 

\begin{itemize}\item Formulae of type $z_1 P_{0,0,m,0}^\kappa (z)$:\beqrn z_1\,P_{0,0,m,0}^\kappa (z)&=& P_{1,0,m,0}^\kappa(z)+b_m(\kappa) 
P_{0,0,m-1,1}(z)\\z_1\,P_{0,0,0,m}^\kappa (z)&=& P_{1,0,0,m}^\kappa(z)+b_m(\kappa) 
P_{0,0,1,m-1}(z)\\z_3\,P_{m,0,0,0}^\kappa (z)&=& P_{m,0,1,0}^\kappa(z)+b_m(\kappa) 
P_{m-1,0,0,1}(z)\\z_3\,P_{0,0,0,m}^\kappa (z)&=& P_{0,0,1,m}^\kappa(z)+b_m(\kappa) 
P_{1,0,0,m-1}(z)\\z_4\,P_{m,0,0,0}^\kappa (z)&=& P_{m,0,0,1}^\kappa(z)+b_m(\kappa) 
P_{m-1,0,1,0}(z)\\z_4\,P_{0,0,m,0}^\kappa (z)&=& P_{0,0,m,1}^\kappa(z)+b_m(\kappa) 
P_{1,0,m-1,0}(z)\\\eeqrn with
\[ b_m(\kappa)=\frac{m(m-1+4\kappa)}{(m-1+\kappa)(m+3\kappa)}\ .\] 
\item Formulae of type $z_1\,P_{0,m,0,0}^\kappa (z)$:\beqrn z_1\,P_{0,m,0,0}^\kappa (z)&=& P_{1,m,0,0}^\kappa(z)+d_m(\kappa) 
P_{1,m-1,0,0}(z)+e_m(\kappa)P_{0,m-1,1,1}^\kappa(z)\\z_3\,P_{0,m,0,0}^\kappa (z)&=& P_{0,m,1,0}^\kappa(z)+d_m(\kappa) 
P_{0,m-1,1,0}(z)+e_m(\kappa)P_{1,m-1,0,1}^\kappa(z)\\z_4\,P_{0,m,0,0}^\kappa (z)&=& P_{0,m,0,1}^\kappa(z)+d_m(\kappa) 
P_{0,m-1,0,1}(z)+e_m(\kappa)P_{1,m-1,1,0}^\kappa(z)\\\eeqrn with\beqrn d_m(\kappa)&=&\frac{2m(m+\kappa)(m-1+3\kappa)(m-1+4\kappa)(2m-1+6\kappa)}
{(m-1+\kappa)(m-1+2\kappa)(m+3\kappa)(2m-1+5\kappa)(2m+5\kappa)},\\e_m(\kappa)&=&\frac{m(m-1+3\kappa)}{(m-1+\kappa)(m+2\kappa)}\ .\eeqrn 
\item Formulae of type $z_2\,P_{m,0,0,0}^\kappa (z)$:\beqrn z_2\,P_{m,0,0,0}^\kappa (z)&=& P_{m,1,0,0}^\kappa(z)+f_m(\kappa)\,
P_{m-2,1,0,0}(z)+g_m(\kappa)\,P_{m-1,0,1,1}^\kappa(z)+h_m(\kappa)\,
P_{m,0,0,0}^\kappa(z)\\z_2\,P_{0,0,m,0}^\kappa (z)&=& P_{0,1,m,0}^\kappa(z)+f_m(\kappa)\,
P_{0,1,m-2,0}(z)+g_m(\kappa)P_{1,0,m-1,1}^\kappa(z)+h_m(\kappa)\,
P_{0,0,m,0}^\kappa(z)\\z_2\,P_{0,0,0,m}^\kappa (z)&=& P_{0,1,0,m}^\kappa(z)+f_m(\kappa)\,
P_{0,1,0,m-2}(z)+g_m(\kappa)\,P_{1,0,1,m-1}^\kappa(z)+h_m(\kappa)\,
P_{0,0,0,m}^\kappa(z)\eeqrn with\beqrn f_m(\kappa)&=&\frac{m(m-1)(m-2+2\kappa)(m+2\kappa)(m-1+4\kappa)(m-1+5\kappa)}
{(m-2+\kappa)(m-1+\kappa)^2(m-1+3\kappa)(m+3\kappa)(m+4\kappa)},\\g_m(\kappa)&=&\frac{m(m-1+3\kappa)}{(m-1+\kappa)(m+2\kappa)},\\h_m(\kappa)&=&\frac{4\left[ -3\kappa^3+5\kappa^2+(6m-1)\kappa+(m^2-1)
\right] }{(m-1+\kappa)(1+3\kappa)(m+1+5\kappa)}\ .\\\eeqrn 
\item Formula for $z_2\,P_{0,m,0,0}^\kappa (z)$:\beqrn z_2\,P_{0,m,0,0}^\kappa (z)&=& P_{0,m+1,0,0}^\kappa(z)+k_m(\kappa)\,
P_{0,m-1,0,0}(z)+p_m(\kappa)\,P_{1,m-1,1,1}^\kappa(z)+ q_m(\kappa)\,
P_{1,m-2,1,1}^\kappa(z)\\
&+&r_m(\kappa)\,\left[P_{2,m-1,0,0}^\kappa(z)+P_{0,m-1,2,0}^\kappa(z)+
P_{0,m-1,0,2}^\kappa(z)\right]+s_m(\kappa)\,P_{0,m,0,0}^\kappa(z)\eeqrn with\beqrn k_m(\kappa)&=&\frac{4 m(m+\kappa)^2(m+2\kappa)(m-1+3\kappa)(m-1+4\kappa)^2
(2m-1+4\kappa)(m-1+5\kappa)(2m-1+6\kappa)}
{(m-1+\kappa)(m-1+2\kappa)^2(m+3\kappa)^2(m+4\kappa)(2m-2+5\kappa)
(2m-1+5\kappa)^2(2m+5\kappa)},\\p_m(\kappa)&=&\frac{m(m-1+2\kappa)}{(m-1+\kappa)(m+\kappa)},\\q_m(\kappa)&=&\frac{2m(m-1)(m+\kappa)^2(m-2+2\kappa)(m-1+3\kappa)^3
(2m-1+6\kappa)}{(m-2+\kappa)(m-1+\kappa)^2(m-1+2\kappa)^2(m+2\kappa)^2
(2m-1+5\kappa)(2m+5\kappa)},\\r_m(\kappa)&=&\frac{m(m+\kappa)(m-1+3\kappa)(m-1+4\kappa)}
{(m-1+\kappa)(m-1+2\kappa)(m+2\kappa)(m+3\kappa)},\\s_m(\kappa)&=&\frac{-4 t_m(\kappa)}{(\kappa+1)(m-1+\kappa)(m+1+4\kappa)
(2m-1+5\kappa)(2m+1+5\kappa)},\\t_m(\kappa)&=&(-1+5m^2-4m^4)+(2+25m-7m^2-40m^3+2m^4)\kappa+
(20-35m-123m^2+20m^3)\kappa^2,\\
&+&(-22-115m+63 m^2)\kappa^3+(-19+65 m)\kappa^4+20\kappa^5.\eeqrn \end{itemize} 

Finally, we mention that for $\kappa=-\frac{1}{2}(n-1)$, $n\in{\bf N}$, the polynomials associated to the dominant weight which is $n$ times the Weyl vector $\rho$ are proportional to a power of the ground state wavefunction, namely
\[
P_{n \rho}^{-\frac{1}{2}(n-1)}= (-1)^n\, 2^{12 n} \left\{ \prod_{j<k}^4\sin(q_j-q_k)\sin(q_j+q_k)
\right\} ^n.
\]
This formula can be verified quite easily by direct application of $\Delta^{-\frac{1}{2}(n-1)}$ in the form (\ref{4b}) to the right-hand side: one finds that the Schr\"{o}dinger equation (\ref{sch}) with the appropriate eigenvalue is satisfied. The most convenient way to fix the proportionality constant is by performing an analytic continuation to complex $q_i$ and considering the region $x_i\in {\bf R}$ and $x_1\gg x_2\gg x_3\gg x_4\gg 0$. Then, the polynomials are dominated by the leading order term, $P_{n \rho}^{-\frac{1}{2}(n-1)}\simeq z_1^n z_2^n z_3^n z_4^n$,  and, on the other hand, using the formulas for the fundamental characters displayed in Section 2. one finds $z_1 z_2 z_3 z_4 \simeq x_1^3 x_2^2 x_3$ and
$\prod_{j<k}^4\sin(q_j-q_k)\sin(q_j+q_k) \simeq -2^{-12}x_1^3 x_2^2 x_3$. This gives the proportionality constant written above.
\section{Conclusions}
In this paper, we have shown how to solve the Schr\"{o}dinger equation 
for the trigonometric Calogero-Sutherland model related to the Lie algebra 
$D_4$ and we have explored some properties of the energy eigenfunctions. 
The main point is that the use of a Weyl-invariant set of variables, 
the characters of the fundamental representations, leads to a formulation 
of the Schr\"odinger equation by means of a second order differential 
operator which is simple enough to make feasible a recursive method for 
the treatment of the spectral problem. The eigenfunctions provide a complete 
system of orthogonal polynomials in four variables, and these polynomials 
obey recurrence relations which are extensions of the Clebsch-Gordan series 
of the algebra. The structure of some of these recurrence relations has been 
fixed and, for particular cases, the coefficients involved have been 
computed. Also, some generating functions for the polynomials with parameter 
$\kappa=1$ and $\kappa=0$ have been obtained. These generating functions can 
give some hints about the form of the generating function for general 
$\kappa$, see \cite{pe00}.\\\\
{\Large\bf Acknowledgements}\\\\
A.M.P. would like to express his gratitude to the Max Planck Institut 
f\"{u}r Mathematik for hospitality. The work of J.F.N. and W.G.F. has been 
partially supported by the University of Oviedo, Vicerrectorado de 
Investigaci\'on, grant MB-03-514-1. 

\section*{Appendix A: Summary of results on the Lie algebra $D_4$}
In this appendix, we review some standard facts about the root and weight 
systems of the Lie algebra $D_4$ that the reader could find useful to follow 
the main text. More extensive and sound treatments of these topics can be 
found in many excellent textbooks, see for instance \cite{ov90}, 
\cite{otros}. 

The most convenient explicit representation of $D_4$ is \bdm D_4=\left\{ \left( \begin{array}{cc}m&b\\c&-m^t\end{array} \right)\, 
\mid\,  m,\,b,\,c\ {\rm real\ }4\times 4\ {\rm matrices\ and\ }
b^t=-b,\ c^t=-c \right\} .\edm This gives $\dim D_4=28$. One can choose the following linear basis:\beqrn M_{jk}&=&E_{j,k}-E_{4+j,4+k},\hspace{1.5cm}j,k=1,2,3,4\\B_{jk}&=&E_{j,4+k}-E_{k,4+j},\hspace{1.5cm}j,k=1,2,3,4,\ \ j<k\\C_{jk}&=&E_{4+j,k}-E_{4+k,j},\hspace{1.5cm}j,k=1,2,3,4,\ \ j<k\eeqrn with $(E_{i,j})_{kl}=\delta_{ik}\delta_{jl}$. The Cartan subalgebra is\bdm H=\left\{ h=\sum_{i=1}^4 c_i M_{ii}\,\mid\,  c_i\in {\bf R} \right\}\edm and this confirms that the rank of $D_4$ is four. The matrix commutators\beqrn \left[h,M_{jk}\right]&=&(c_j-c_k)M_{jk},\\\left[ h,B_{jk}\right]&=&(c_j+c_k)B_{jk},\\\left[h,C_{jk}\right]&=&-(c_j+c_k)C_{jk}\eeqrn allow us to classify the 24 roots in two groups\beqrn \alpha_{jk}(h)&=&c_j-c_k,\hspace{1.5cm} j\neq k,\\\alpha_{jk}^\pm(h)&=&\pm (c_j+c_k),\hspace{1.2cm} j < k .\eeqrn 
One can extract the following basis of simple roots\beqrn \alpha_1\equiv \left(\begin{array}{cccc}1,&-1,&0,&0\end{array}\right) 
=\alpha_{12},\hspace{1.5cm}\alpha_2\equiv\left(\begin{array}{cccc}0,&1,&-1,&0
\end{array}\right)=\alpha_{23},\\\alpha_3\equiv\left(\begin{array}{cccc}0,&0,&1,&-1\end{array}\right)=
\alpha_{34},\hspace{1.5cm}\alpha_4\equiv\left(\begin{array}{cccc}0,&0,&1,&1
\end{array}\right)=\alpha_{34}^+,\eeqrn where we have given the decomposition of these roots in the basis of $H^*$ 
dual to ${\rm diag}(M_{ii})$, i=1,2,3,4. The euclidean relations among 
the simple roots are\beqrn (\alpha_i,\alpha_i)&=&2,\hspace{1.5cm} i=1,2,3,4,\\(\alpha_2,\alpha_i)&=&-1,\hspace{1.2cm} i=1,3,4,\\(\alpha_i,\alpha_j)&=&0,\hspace{1.5cm} i=1,3,4.\eeqrn Thus, the Cartan matrix reads\bdm A=\left(\begin{array}{cccc}2&-1&0&0\\-1&2&-1&-1\\0&-1&2&0\\0&-1&0&2 \end{array}\right) .
\edm The positive roots are $\alpha_{ij},\alpha_{ij}^+, i<j$, and they can be 
classified by heights as indicated in the table.\begin{table}[h]\begin{center}\begin{tabular}{|c|l|}\hline {\rm Height} & {\rm Positive roots} \\\hline 1 & $\alpha_1,\ \alpha_2,\ \alpha_3,\ \alpha_4$ \\\hline 2 & $\alpha_{13}=\alpha_1+\alpha_2,\  \alpha_{24}=\alpha_2+\alpha_3,\ 
\alpha_{24}^+=\alpha_2+\alpha_4 $\\\hline 3 & $\alpha_{14}=\alpha_1+\alpha_2+\alpha_3,\ \alpha_{14}^+=
\alpha_1+\alpha_2+\alpha_4,\ \alpha_{23}^+=\alpha_2+\alpha_3+\alpha_4$\\\hline 4 & $\alpha_{13}^+=\alpha_1+\alpha_2+\alpha_3+\alpha_4$ \\\hline 5 & $\alpha_{12}^+=\alpha_1+2\alpha_2+\alpha_3+\alpha_4$ \\\hline \end{tabular}\caption[smallcaption]{Heights of positive roots.}\end{center}\end{table}
The Weyl group is easy to describe. The Weyl reflection on the hyperplane in 
$H^*$ orthogonal to the root $\alpha$ is $s_\alpha(v)=
v-2\frac{(\alpha,v)}{(\alpha,\alpha)}\alpha$. Applying this formula to 
$\alpha_{ij}, \alpha_{ij}^\pm$, one readily finds that the most general Weyl 
reflection consists in a permutation of the components of $v$ in the $e_i$ 
basis plus an even number of changes of the signs of these components. 
This gives $|W|=192$ for the order of the Weyl group.
The fundamental weights $\lambda_k$ can be obtained from the equation 
$\alpha_i=\sum_{j=1}^4 A_{ji}\lambda_j$. They are\beqrn \lambda_1&=&\frac{1}{2}(2\alpha_1+2\alpha_2 +\alpha_3 +\alpha_4)=
\frac{1}{2}\;(\begin{array}{cccc}2,&0,&0,&0\end{array}),\\\lambda_2&=&\frac{1}{2}(2\alpha_1+4\alpha_2 +2\alpha_3 +2\alpha_4)=
\frac{1}{2}\;(\begin{array}{cccc}2,&2,&0,&0\end{array}),\\\lambda_3&=&\frac{1}{2}(\alpha_1+2\alpha_2 +2\alpha_3 +\alpha_4)=
\frac{1}{2}\;(\begin{array}{cccc}1,&1,&1,&-1\end{array}),\\\lambda_4&=&\frac{1}{2}(\alpha_1+2\alpha_2 +\alpha_3 +2\alpha_4)=
\frac{1}{2}\;(\begin{array}{cccc}1,&1,&1,&1\end{array}),\eeqrn and the geometry of the weight system is summarized by the relations\beqrn \parallel \lambda_1 \parallel = \parallel \lambda_3 \parallel = 
\parallel \lambda_4 \parallel =1,&\ \ \ \ \ &\parallel \lambda_2 
\parallel =\sqrt{2}, \\(\lambda_i ,\lambda_2)=1,\ \ \ i=1,3,4,\ \ &\ \ \ \ \ &(\lambda_i ,
\lambda_j)=\frac{1}{2},\ \ i,j=1,3,4 .\eeqrn The Weyl vector is\bdm \rho =\frac{1}{2}\sum_{\alpha\in R^+}=\sum_{j=1}^4 \lambda_j=\alpha=3\alpha_1 +5\alpha_2 +3\alpha_3 +
3\alpha_4=\,(\begin{array}{cccc}3,&2,&1,&0\end{array}),\edm and the Weyl formula for dimensions applied to the irreducible representation 
associated to the integral dominant weight ${\bf m}=m_1\lambda_1+
m_2\lambda_2+m_3\lambda_3+m_4\lambda_4$ gives\bdm \dim r({\bf m})=\prod_{\alpha\in R^+} \frac{(\alpha, {\bf m}+\rho)}
{(\alpha,\rho)}=\frac{P}{1440}\edm with\bdm P=\prod_{i=1}^4(m_i+1)\;\prod_j(m_2+m_j+2)\;\prod_{j<k}(m_2+m_j+m_k+3)\;
(m_1+m_2+m_3+m_4)\;(m_1+2m_2+m_3+m_4)\edm where the indices $j,k$ take the values $1,3,4$. In particular, 
for the fundamental representations, one finds:\beqrn \dim r(\lambda_1)=8,&\ \ \ \ \ \ \ \ \ &\dim r(\lambda_2)=28,\\\dim r(\lambda_3)=8,&\ \ \ \ \ \ \ \ \ &\dim r(\lambda_4)=8.\eeqrn 

\section*{Appendix B: Some polynomials, characters and monomial functions}
We list here all the polynomials, characters and monomial 
functions with total degree lower or equal to three up to triality. 

\subsection*{Polynomials}
\beqrn P_{1,0,0,0}^\kappa(z)&=&z_1\\
P_{0,1,0,0}^\kappa(z)&=&z_2+\frac{4(\kappa-1)}{5\kappa+1},\\
P_{2,0,0,0}^\kappa(z)&=&z_1^2-\frac{2}{1+\kappa}\,z_2-\frac{8\kappa}
{(1+\kappa)(1+3\kappa)}\\
P_{0,2,0,0}^\kappa(z)&=&z_2^2-\frac{2}{1+\kappa}\,z_1z_3z_4-
\frac{2(-1+\kappa)}{(1+\kappa)(1+2\kappa)}\,(z_1^2+z_3^2+z_4^2)+
\frac{4(-3+5\kappa+6\kappa^2+4\kappa^3)}{(1+\kappa)(1+2\kappa)(3+5\kappa)}\,
z_2+\\
&+& \frac{16(-1+\kappa)(3+10\kappa+3\kappa^2+2\kappa^3)}
{(1+\kappa)(1+2\kappa)(2+5\kappa)(3+5\kappa)}\\
P_{1,1,0,0}^\kappa(z)&=&z_1 z_2-\frac{3}{1+2\kappa}\,z_3z_4+
\frac{4(-1+\kappa)(-1+2\kappa)}{(1+2\kappa)(2+5\kappa)}z_1\\
P_{1,0,1,0}^\kappa(z)&=&z_1 z_3-\frac{4}{1+3\kappa}z_4\\P_{3,0,0,0}^\kappa(z)&=&z_1^3-\frac{6}{2+\kappa}z_1 z_2+\frac{6}{(1+\kappa)
(2+\kappa)}z_3 z_4-\frac{12(1+2\kappa+2\kappa^2)}{(1+\kappa)(2+\kappa)
(2+3\kappa)}\,z_1\\
P_{0,3,0,0}^\kappa(z)&=&z_2^3-\frac{6}{2+\kappa}\,z_1 z_2 z_3z_4+
\frac{6}{(1+\kappa)(2+\kappa)}\,(z_1^2z_3^2+z_1^2z_4^2+z_3^2z_4^2)-
\frac{3(2+\kappa+\kappa^2)}{(1+\kappa )^2(2+\kappa)}\,(z_1^2 z_2+z_2 z_3^2+
z_2 z_4^2)\\
&+&\frac{6(10+17\kappa+21\kappa^2+10\kappa^3+2\kappa^4)}
{5(1+\kappa )^3(2+\kappa)}\,z_2^2-\frac{3(30+53\kappa+4\kappa^2-15\kappa^3+
8\kappa^4)}{5(1+\kappa)^4(2+\kappa)}\,z_1 z_3 z_4\\
&-&\frac{12\kappa(8+10\kappa+\kappa^2+\kappa^3)}{5(1+\kappa)^4(2+\kappa)}
(z_1^2+z_3^2+z_4^2)+\frac{12(30+119\kappa+159\kappa^2+124\kappa^3+
80\kappa^4+24\kappa^5+4\kappa^6)}{5(1+\kappa)^4(2+\kappa)(4+5\kappa)}z_2\\
&+&\frac{16(-30+103\kappa+440\kappa^2+359\kappa^3+98\kappa^4+86\kappa^5+
20\kappa^6+4\kappa^7)}{5(1+\kappa)^4(2+\kappa)(3+5\kappa)(4+5\kappa)}\\
P_{2,1,0,0}^\kappa(z)&=&z_1^2z_2-\frac{2}{1+\kappa}z_2^2-\frac{1+3\kappa}
{(1+\kappa)^2}\,z_1 z_3 z_4+\frac{4(-1+\kappa)\kappa^2}{(1+\kappa)^2
(3+5\kappa)}\,z_1^2+\frac{4}{(1+\kappa)^2}(z_3^2+z_4^2)\\
&-&\frac{4(9+27\kappa+28\kappa^2+16\kappa^3)}{(1+\kappa)^2(2+3\kappa)
(3+5\kappa)}z_2-\frac{16(3+5\kappa+2\kappa^3)}{(1+\kappa)^2(2+3\kappa)
(3+5\kappa)}\\
P_{1,2,0,0}^\kappa(z)&=&z_1 z_2^2-\frac{2}{1+\kappa}z_1^2z_3 z_4-
\frac{1+3\kappa}{(1+\kappa)^2}\,z_2 z_3 z_4-\frac{2(-1+\kappa)}{(1+\kappa)
(1+2\kappa)}z_1^3+\frac{5-\kappa}{(1+\kappa)^2}(z_1 z_3^2+z_1 z_4^2)\\
&+&\frac{4(-1+\kappa)(9+19\kappa+10\kappa^2+4\kappa^3)}{(1+\kappa)^2
(1+2\kappa)(4+5\kappa)}z_1 z_2-\frac{4(-1+\kappa)(-5+2\kappa)(1+3\kappa)}
{(1+\kappa)^2(1+2\kappa)(4+5\kappa)}z_3 z_4\\
&+&\frac{8(-9-57\kappa-72\kappa^2+28\kappa^3-2\kappa^4+4\kappa^5)}
{(1+\kappa)^2(1+2\kappa)(3+5\kappa)(4+5\kappa)}\,z_1\\
P_{1,1,1,0}^\kappa(z)
&=&z_1 z_2 z_3-\frac{3}{1+2\kappa}(z_1^2 z_4+z_3^2z_4)-
\frac{8(-1+\kappa)}{(1+2\kappa)(2+3\kappa)}z_2 z_4+\frac{4(12+23\kappa-
11\kappa^2+6\kappa^3)}{(1+2\kappa)(2+3\kappa)(3+5\kappa)}z_1 z_3\\
&-&\frac{8(3-22\kappa+4\kappa^2)}{(1+2\kappa)(2+3\kappa)(3+5\kappa)}z_4\\
P_{1,0,1,1}^\kappa(z)
&=&z_1 z_3 z_4-\frac{4}{1+3\kappa}(z_1^2+z_3^2+z_4^2)+
\frac{12}{(1+2\kappa)(1+3\kappa)}z_2+\frac{16(1+5\kappa)}{(1+2\kappa)
(1+3\kappa)^2}\eeqrn 

\subsection*{Characters}
\beqrn P_{1,0,0,0}^{(1)}(z)&=&z_1\\P_{0,1,0,0}^{(1)}(z)&=&z_2\\P_{2,0,0,0}^{(1)}(z)&=&z_1^2-z_2-1\\P_{0,2,0,0}^{(1)}(z)&=&z_2^2+z_2-z_1 z_3 z_4\\P_{1,1,0,0}^{(1)}(z)&=&z_1 z_2-z_3 z_4\\P_{1,0,1,0}^{(1)}(z)&=&z_1 z_3-z_4\\P_{3,0,0,0}^{(1)}(z)&=&z_1^3-2z_1 z_2+z_3 z_4-2z_1\\P_{0,3,0,0}^{(1)}(z)&=&z_2^3+3z_2^2+3z_2-2z_1 z_2z_3z_4+z_1^2 z_3^2+
z_1^2 z_4^2+z_3^2z_4^2\\
&-&(z_1^2+z_3^2+z_4^2)z_2-z_1 z_3 z_4-z_1^2-z_3^2-z_4^2+1\\P_{2,1,0,0}^{(1)}(z)&=&z_1^2 z_2-z_2^2-z_1 z_3 z_4+z_3^2+z_4^2-2z_2-1\\P_{1,2,0,0}^{(1)}(z)&=&z_1 z_2^2-z_1^2z_3z_4-z_2 z_3 z_4+z_1(z_3^2+z_4^2)-
z_1\\P_{1,1,1,0}^{(1)}(z)&=&z_1 z_2 z_3+z_1 z_3-(z_1^2 +z_3^2)z_4+z_4\\P_{1,0,1,1}^{(1)}(z)&=&z_1 z_3 z_4-z_1^2-z_3^2-z_4^2+z_2+2\eeqrn 
\subsection*{Monomial functions}
\beqrn P_{1,0,0,0}^{(0)}(z)&=&z_1\\P_{0,1,0,0}^{(0)}(z)&=&z_2-4\\P_{2,0,0,0}^{(0)}(z)&=&z_1^2-2 z_2\\P_{0,2,0,0}^{(0)}(z)&=&z_2^2-2\,z_1 z_3 z_4+2\,z_1^2+2\,z_3^2+2\,z_4^2-4\,z_2
-8\\P_{1,1,0,0}^{(0)}(z)&=&z_1 z_2-3z_3 z_4+2z_1\\P_{1,0,1,0}^{(0)}(z)&=&z_1 z_3-4z_4\\P_{3,0,0,0}^{(0)}(z)&=&z_1^3-3z_1 z_2+3z_3 z_4-3z_1\\P_{0,3,0,0}^{(0)}(z)&=&z_2^3+6z_2^2+9z_2-3z_1 z_2z_3z_4+3z_1^2 z_3^2+
3z_1^2 z_4^2+3z_3^2z_4^2\\&-&3(z_1^2+z_3^2+z_4^2)z_2-9z_1 z_3 z_4-4\\P_{2,1,0,0}^{(0)}(z)&=&z_1^2 z_2-2z_2^2-z_1 z_3 z_4+4z_3^2+4z_4^2-6z_2-8\\P_{1,2,0,0}^{(0)}(z)&=&z_1 z_2^2-2z_1^2z_3z_4-z_2 z_3 z_4+2 z_1^3 +
5z_1(z_3^2+z_4^2)-9z_1 z_2-5z_3 z_4-6z_1\\P_{1,1,1,0}^{(0)}(z)&=&z_1 z_2 z_3+8z_1 z_3-3(z_1^2+z_3^2)z_4+4z_2z_4-4z_4\\P_{1,0,1,1}^{(0)}(z)&=&z_1 z_3 z_4-4z_1^2-4z_3^2-4z_4^2+12 z_2+16\eeqrn

\end{document}